\documentstyle[12pt]{article}
\def\text#1{\mbox{$#1$}}

\begin{document}
\title{\bf The Lax pair for $C_{2}$-type Ruijsenaars-Schneider
model\thanks{Project supported by the National Natural
Science Foundation of China(Grant No. 19805006)}}
\author{Kai Chen, Bo-yu Hou and Wen-li Yang\\
{\small {\it Institute of Modern Physics, Northwest University, Xian 710069,
China }}
}

\maketitle

\begin{abstract}
We study the $C_{2}$ Ruijsenaars-Schneider(RS) model with interaction
potential of  trigonometric type. The Lax pairs for the
model with and without spectral parameter are constructed.
Also given are the involutive
Hamiltonians for the system. Taking nonrelativistic limit, we obtain
the Lax pair of $C_{2}$	 Calogero-Moser model.\newline
{\bf \noindent PACC:} 7520H; 1240E; 7510D
\end{abstract}

\section{Introduction}

\noindent Ruijsenaars-Schneider(RS) and Calogero-Moser(CM) models as
integrable many-body models recently have attracted remarkable attention and
have been extensively studied. They describe one-dimensional
$n$-particle system with pairwise interaction. Their importance lies in
various fields ranging from lattice models in statistical physics$^{\small \cite{h1, nksr}}$,
 to the field theory and  gauge theory$^{\small \cite{gm, n}}$, e.g., to the Seiberg-Witten theory
$^{\small \cite{bmmm}}$, etc. Recently, the Lax pairs for the  CM models in various
root systems have been given by Olshanetsky {\sl et al}$^{\small \cite{op}}$, Inozemtsev$^{\cite{in}}$,
D'Hoker {\sl et al}$^{\small \cite{hp1}}$ and  Bordner {\sl et al}$^{\small \cite{bcs}}$
with or without spectral parameter respectively. Further
a more general algebra-geometric construction was proposed by Hurtubise {\sl et al}
in Ref. \cite{hm}, while the commutative
operators for the RS model based on various type Lie algebra
were given by Komori$^{\small \cite{ko1, ko2}}$, Diejen$^{\small \cite{di, di1}}$ and Hasegawa
{\sl et al}$^{\small \cite{h1, h2}}$. An interesting
result is that in Ref. \cite{kai1}, the authors show that for the $sl_{2}$
trigonometric RS and CM models there exists the same non-dynamical $r$-matrix
structure compared with the usual dynamical ones. On the other hand, similar
to Hasegawa's result that $A_{N-1}$  RS model can be obtained as transfer
matrices associated to the Sklyanin algebra, they also reveal that
corresponding CM model's integrability can be depicted by $sl_{N}$ Gaudin
algebra$^{\cite{kai2}}$.

As for the $C_{n}$ type RS model, commuting difference operators acting on
the space of functions on the $C_{2}$ type weight space have been
constructed by Hasegawa {\sl et al} in Ref. \cite{h2}. Extending that work, the
diagonalization of elliptic difference system of that type has been studied
by Kikuchi in Ref. \cite{ki}. Despite of the fact that the Lax pairs for CM
models have been proposed for general Lie algebra even for all of the finite
reflection groups$^{\small \cite{bcs1}}$, the Lax integrability of RS model are
not clear except only for $A_{N-1}$ -type$^{\small \cite{r1, nksr, bc, kz, s1, s2}}$, i.e.,
the Lax pairs for the RS models other than $A_{N-1}$ -type have not yet
been obtained.

In this paper, we concentrate on  the $C_2$ type trigonometric
Ruijsenaars-Schneider model. The basic materials about $C_{2}$ $RS$
model are reviewed in Section \ref{model}. In Section \ref{pair}, we  present the Lax
pair without spectral parameter	 and its integrability in Liouville
sense is also given. In Section \ref{limit}, taking its non-relativistic limit, we
recover the system of corresponding CM type. In Section \ref{spectral},
we give the Lax pair for the system  with spectral parameter, and show that at certain	limit it
will degenerate to the one without spectral parameter. The last
section is a brief summary and some discussions.

\section{Model and equations of motion}
\label{model}
As a relativistic-invariant generalization of the $C_{n}$-type
Calogero-Moser model, the $C_{n}$-type Ruijsenaars-Schneider
system is  completely integrable  whose integrability is
shown by Ruijsenaars$^{\small \cite{r2}}$ and  Diejen
$^{\small \cite{di, di1}}$. In terms of the canonical
variables $p_{i}$, $x_{i} (i,j=1, \ldots, 2)$ enjoying in the canonical
Poisson bracket $\{p_{i},p_{j}\}=\{x_{i},x_{j}\}=0\,,$ \ $%
\{x_{i},p_{j}\}=\delta _{ij}\,$,  the Hamiltonian of $C_{2}$ RS system
can be of the form
\begin{eqnarray}
H &=&\sum_{i=1}^{2}\{e^{p_{i}}f(2x_{i})\,\prod_{k\neq
i}^2(f(x_{ik})f(x_{i}+x_{k}))  \nonumber \\
&&+e^{-p_{i}}g(2x_{i})\,\prod_{k\neq i}^2(g(x_{ik})g(x_{i}+x_{k}))\,\},
\label{hami}
\end{eqnarray}
where
\begin{eqnarray*}
f(x) &:&=\frac{\sin (x-\gamma )}{\sin (x)},\\
g(x) &:&=f(x)|_{\gamma \rightarrow -\gamma },~~~~~~~~~~~~~
x_{ik}:=x_{i}-x_{k},
\end{eqnarray*}
and $\gamma $ denotes the coupling constant. Notice that  in
Ref. \cite{r2} Ruijsenaars used another ``gauge" of the
momenta such that the two systems are
connected by the following canonical transformation:

\begin{eqnarray}
x_{i}\longrightarrow x_{i},\ \ \ \ p_{i}\longrightarrow p_{i}+\frac{1}{2}%
ln\prod_{j\neq i}^{2}\frac{f(x_{ij})f(x_{i}+x_{j})}{g(x_{ij})g(x_{i}+x_{j})}%
\frac{f(2x_{i})}{g(2x_{i})}.
\end{eqnarray}

The canonical equations of motion for the Hamiltonian (\ref{hami}) are
\begin{eqnarray}
\dot{x_{i}} &=&\{x_{i},H\}=e^{p_{i}}b_{i}-e^{-p_{i}}b_{i}^{^{\prime }},
\label{equ1} \\
\dot{p_{i}} &=&\{p_{i},H\}=\sum_{j\neq i}^{2}\Big(e^{p_{j}}b_{j}\big(%
h(x_{ji})-h(x_{j}+x_{i})\big)  \nonumber \\
&&+e^{-p_{j}}b_{j}^{^{\prime }}\big(\hat{h}(x_{ji})-\hat{h}(x_{j}+x_{i})\big)%
\Big)  \nonumber \\
&&-e^{p_{i}}b_{i}\Big(2h(2x_{i})+\sum_{j\neq i}^{2}\big(%
h(x_{ij})+h(x_{i}+x_{j})\big)\Big)  \nonumber \\
&&-e^{-p_{i}}b_{i}^{^{\prime }}\Big(2\hat{h}(2x_{i})+\sum_{j\neq i}^{2}\big(%
\hat{h}(x_{ij})+\hat{h}(x_{i}+x_{j})\big)\Big),	  \label{equ2}
\end{eqnarray}
where
\begin{eqnarray}
h(x):&=&\frac{d\ln f(x)}{dx},\text{ \ \ \ \ \ \ \ \ \ \ \ }\hat{h}(x):=\frac{%
d\ln g(x)}{dx},	 \nonumber \\
b_{i} &=&f(2x_{i})\,\prod_{k\neq i}^{2}\Big(f(x_{i}-x_{k})f(x_{i}+x_{k})\Big),
\nonumber \\
b_{i}^{^{\prime }} &=&g(2x_{i})\,\prod_{k\neq
i}^{2}\Big(g(x_{i}-x_{k})g(x_{i}+x_{k})\Big).
\end{eqnarray}
Here,\vspace{1pt} of course $x_{i}=x_{i}(t)$,  $p_{i}=p_{i}(t)$	 and the
dot on	top denotes t-differentiation.

\section{The Lax pair without spectral parameter}
\label{pair}
Let us first mention some results about the integrability of
Hamiltonian (\ref{hami}). In Ref. \cite{r2} Ruijsenaars
demonstrated that the symplectic structure of $C_{n}$ type
RS system can be proved integrable by embedding its phase
space to a submanifold of $A_{2n-1}$ type RS one, while in
Refs. \cite{di,di1} and Ref. \cite{ko2}, Diejen and Komori, respectively,  gave a
series of commuting difference operators which led to its
quantum integrability. However, there is not any result
about its Lax representation so far. That is, the explicit
form of the Lax matrix $L$, associated with a $M$ which
ensure its Lax integrability,  have not been proposed up to
now. In this section, we restrict our treatment to the
exhibition of the explicit form for $C_{2}$  RS system.
Therefore, some previous results, as well as new results,
could now be obtained in a more straightforward manner by
using the Lax pair.

Define one $4\times 4$ Lax matrix for $C_{2}$ RS model as follows:

\begin{eqnarray}
L=\left(
\begin{array}{ll}
A & B \\
C & D
\end{array}
\right),  \label{lax}
\end{eqnarray}
where $A$, $B$, $C$, $D$ are $2\times 2$ matrices(hereafter, we use the indices $i, j=1,2$)

\begin{eqnarray}
A_{ij} &=&e^{p_{j}}b_{j}\frac{\sin \gamma }{\sin (x_{ij}+\gamma )},\ \ \ \ \
\ \ \ \ \ \ B_{ij}=e^{-p_{j}}b_{j}^{^{\prime }}\frac{\sin \gamma }{\sin
(x_{i}+x_{j}+\gamma )},	 \nonumber \\
\ C_{ij} &=&e^{p_{j}}b_{j}\frac{\sin \gamma }{\sin (-x_{i}-x_{j}+\gamma )},\
\ \ D_{ij}=e^{-p_{j}}b_{j}^{^{\prime }}\frac{\sin \gamma }{\sin
(x_{ji}+\gamma )}. \label{l1}
\end{eqnarray}

For the concise expression for $M$,   we define four auxiliary	$2\times 2$ matrices
$\widetilde{A}$, $\widetilde{B}$, $\widetilde{C}$, $\widetilde{D}$ as follows

\begin{eqnarray}
\widetilde{A}_{ij} &=&e^{-p_{i}}b_{j}^{^{\prime }}\frac{-\sin \gamma }{\sin
(x_{ij}-\gamma )},\ \ \ \ \ \ \ \ \ \ \ \widetilde{B}_{ij}=e^{-p_{i}}b_{j}\;%
\frac{-\sin \gamma }{\sin (x_{i}+x_{j}-\gamma )},  \nonumber \\
\widetilde{C}_{ij} &=&e^{p_{i}}b_{j}^{^{\prime }}\frac{-\sin \gamma }{\sin
(-x_{i}-x_{j}-\gamma )},\ \ \ \ \ \widetilde{D}_{ij}=e^{p_{i}}b_{j}\frac{%
-\sin \gamma }{\sin (x_{ji}-\gamma )},\label{lax1}
\end{eqnarray}
such that $M$ can be  of the form

\begin{eqnarray}
M=\left(
\begin{array}{ll}
\cal{A} & \cal{B} \\
\cal{C} & \cal{D}
\end{array}
\right),
\end{eqnarray}
where entries of $M$ are

\begin{eqnarray}
{\cal A}_{ij} &=&\cot (x_{ij})(A_{ij}-\widetilde{A}_{ij}),\ \ \ \ \ \ \ \ \ \ \
{\cal D}_{ij}=\cot (x_{ji})(D_{ij}-\widetilde{D}_{ij}),\ \ \ \ \ \ \ \ \ (i\neq j),\nonumber \\
{\cal B}_{ij}&=&\cot (x_{i}+x_{j})(B_{ij}-\ \widetilde{B}_{ij}),\ \ \ \ \
{\cal C}_{ij} =\cot (-x_{i}-x_{j})(C_{ij}-\widetilde{C}_{ij}),\nonumber \\
{\cal A}_{ii} &=&-\sum_{k\neq i}^{2}\frac{A_{ik}-\widetilde{A}_{ik}}{\sin (x_{ik})}%
-\sum_{k=1}^{2}\frac{B_{ik}-\ \widetilde{B}_{ik}}{\sin (x_{i}+x_{k})},
\nonumber \\
{\cal D}_{ii} &=&\sum_{k\neq i}^{2}\frac{D_{ik}-\widetilde{D}_{ik}}{\sin (x_{ik})}%
+\sum_{k=1}^{2}\frac{C_{ik}-\ \widetilde{C}_{ik}}{\sin
(x_{i}+x_{k})}.\label{me1}
\end{eqnarray}

We have checked that $L, M$ satisfies the Lax equation

\begin{eqnarray}
\dot{L}=\{L,H\}=\lbrack M,L\rbrack,
\label{laxeq}
\end{eqnarray}
which is equivalent to the equations of motion (\ref{equ1}) and (\ref{equ2})
with the help of  computer. The Hamiltonian $H$ can be rewritten in the following form

\begin{eqnarray}
H=\sum_{j=1}^{2}(e^{p_{j}}b_{j}+e^{-p_{j}}b_{j}^{^{\prime }})=tr L.
\label{Ham}
\end{eqnarray}

\vspace{1pt}
The characteristic polynomial of the Lax matrix $L$ is
\begin{eqnarray}
\det (L-v\cdot Id) &=&\sum_{j=0}^{4}(-v)^{4-j}H_{j}  \nonumber \\
&=&v^{4}-H v^{3}+H_{2} v^{2}-H v+1,
\end{eqnarray}
where $H_{0}=H_{4}=1$,	$H_{1}=H_{3}=H$. The function-independent
Hamiltonian flows
$H$ and $H_{2}$ are
\begin{eqnarray}
H &=&e^{p_{1}}f(2x_{1})\,f(x_{12})f(x_{1}+x_{2})  \nonumber \\
&&+e^{-p_{1}}g(2x_{1})\,g(x_{12})g(x_{1}+x_{2})	 \nonumber \\
&&+e^{p_{2}}f(2x_{2})\,f(x_{21})f(x_{2}+x_{1})	\nonumber \\
&&+e^{-p_{2}}g(2x_{2})\,g(x_{21})g(x_{2}+x_{1}),
\label{h1} \\
H_{2} &=&e^{p_{1}+p_{2}}f(2x_{1})\,(f(x_{1}+x_{2}))^{2}f(2x_{2}) \nonumber\\
&&+e^{-p_{1}-p_{2}}g(2x_{1})\,(g(x_{1}+x_{2}))^{2}g(2x_{2})  \nonumber \\
&&+e^{p_{1}-p_{2}}f(2x_{1})\,(f(x_{12}))^{2}f(-2x_{2})	\nonumber \\
&&+e^{p_{2}-p_{1}}g(2x_{1})\,(g(x_{12}))^{2}g(-2x_{2})	\nonumber \\
&&+2f(x_{12})\,g(x_{12})\,f(x_{1}+x_{2})g(x_{1}+x_{2}).
\label{h2}
\end{eqnarray}
We verify  that $H$ and $H_2$  Poisson commute each other
\begin{eqnarray}
\{H,H_{2}\}=0,
\end{eqnarray}
which  ensures the complete integrability of $C_2$ RS model (in Liouville
sense).

\section{Nonrelativistic limit to the Calogero-Moser system}
\label{limit}
The Nonrelativistic limit can be achieved by rescaling \ $%
p_{i}\longmapsto \beta p_{i}$, $\gamma \longmapsto \beta \gamma $ while
letting $\beta \longmapsto 0,$ and making a canonical transformation

\begin{eqnarray}
p_{i}\longmapsto p_{i}+\gamma \left(\cot (2x_{i})+\sum_{k\neq i}^{2}\left(\cot
(x_{ik})+\cot (x_{i}+x_{k})\right)\right),
\end{eqnarray}
such that
\begin{eqnarray}
L &\longmapsto& Id+\beta L_{CM}+O(\beta ^{2}),\\
M &\longmapsto& 2 \beta M_{CM}+O(\beta ^{2}),
\end{eqnarray}
and

\begin{eqnarray}
H &\longmapsto& 4+2 \beta^{2} H_{CM}+O(\beta ^{2}).
\end{eqnarray}
$L_{CM}$ can be expressed as
\begin{eqnarray}
L_{CM}=\left(
\begin{array}{ll}
A_{CM} & B_{CM} \\
-B_{CM} & -A_{CM}
\end{array}
\right),\label{lcm1}
\end{eqnarray}
where

\begin{eqnarray}
(A_{CM})_{ii}&=&p_{i},\ \ \ \ \ \ \ (B_{CM})_{ij}=\frac{\gamma }{\sin (x_{i}+x_{j})},\nonumber\\
(A_{CM})_{ij}&=&\frac{\gamma }{\sin (x_{ij})}, \ \ \ \	(i\neq
j).
\end{eqnarray}
$M_{CM}$ is
\begin{eqnarray}
M_{CM}=\left(
\begin{array}{ll}
{\cal A}_{CM} & {\cal B}_{CM} \\
{\cal B}_{CM} & {\cal A}_{CM}
\end{array}
\right),
\end{eqnarray}
where

\begin{eqnarray}
({\cal A}_{CM})_{ii} &=&-\sum_{k\neq i}^{2}(\frac{\gamma }{\sin ^{2}x_{ik}}+\frac{\gamma
}{\sin ^{2}(x_{i}+x_{k})})-\frac{\gamma }{\sin ^{2}(2x_{i})},\ \ \ \
({\cal B}_{CM})_{ij}=\frac{\gamma \cos (x_{i}+x_{j})}{\sin ^{2}(x_{i}+x_{j})}, \nonumber \\
({\cal A}_{CM})_{ij} &=&\frac{\gamma \cos (x_{ij})}{\sin ^{2}x_{ij}},\ \ \ \ \ (\ i\neq j),
\label{mcm1}
\end{eqnarray}
which coincide with the form given in  Ref. \cite{op} with the difference of a
constant diagonalized matrix.

The Hamiltonian of $C_{2}$-type CM model can be given by

\begin{eqnarray}
H_{CM} &=&\frac{1}{2}\sum_{k=1}^{2}p_{k}^{2}-\frac{\gamma^2 }{2}\sum_{k\neq
i}^{2}(\frac{1}{\sin ^{2}x_{ik}}+\frac{1}{\sin ^{2}(x_{i}+x_{k})}+
\frac{1}{\sin ^{2}(2x_{i})})  \nonumber \\
&=&\frac{1}{4}trL^{2}.
\end{eqnarray}
The $L_{CM}$, $M_{CM}$ satisfies the Lax equation

\begin{eqnarray}
\dot{L}_{CM}=\{L_{CM},H_{CM}\}=\lbrack M_{CM},L_{CM}\rbrack.
\end{eqnarray}

\section{The Lax pair with spectral parameter}
\label{spectral}
Also, we can give the Lax pair which include spectral parameters.
Define the Lax matrix for  trigonometric RS model    as follows:

\begin{eqnarray}
L=\left(
\begin{array}{ll}
A & B \\
C & D
\end{array}
\right),
\label{l2}
\end{eqnarray}
where $A$, $B$, $C$, $D$ are $2\times 2$ matrices($i,j=1,2$)

\begin{eqnarray}
A_{ij} &=&e^{p_{j}}b_{j}\frac{\sin (x_{ij}+\gamma +\lambda )\sin \gamma }{%
\sin (x_{ij}+\gamma )\sin (\gamma +\lambda )},\ \ \ \ \ \ \ \ \ \ \
B_{ij}=e^{-p_{j}}b_{j}^{^{\prime }}\frac{\sin (x_{i}+x_{j}+\gamma +\lambda
)\sin \gamma }{\sin (x_{i}+x_{j}+\gamma )\sin (\gamma +\lambda )},  \nonumber
\\
\ C_{ij} &=&e^{p_{j}}b_{j}\frac{\sin (-x_{i}-x_{j}+\gamma +\lambda )\sin
\gamma }{\sin (-x_{i}-x_{j}+\gamma )\sin (\gamma +\lambda )},\ \ \
D_{ij}=e^{-p_{j}}b_{j}^{^{\prime }}\frac{\sin (x_{ji}+\gamma +\lambda )\sin
\gamma }{\sin (x_{ji}+\gamma )\sin (\gamma +\lambda )}.
\end{eqnarray}

\noindent $M$ is

\begin{eqnarray}
M=\left(
\begin{array}{ll}
\mathcal{A} & \mathcal{B} \\
\mathcal{C} & \mathcal{D}
\end{array}
\right),
\end{eqnarray}
where entries of $M$ are

\begin{eqnarray}
{\mathcal{A}}_{ij} &=&e^{p_{j}}b_{j}\frac{\sin (x_{ij}+\lambda )}{\sin \lambda
\sin x_{ij}}-e^{-p_{i}}b_{j}^{^{\prime }}\frac{\sin (x_{ij}+\lambda +4\gamma
)}{\sin (\lambda +4\gamma )\sin x_{ij}},  \nonumber \\
\ \ \ {\mathcal{D}}_{ij} &=&e^{-p_{j}}b_{j}^{^{\prime }}\frac{\sin
(x_{ji}+\lambda )}{\sin \lambda \sin x_{ji}}-e^{p_{i}}b_{j}\frac{\sin
(x_{ji}+\lambda +4\gamma )}{\sin (\lambda +4\gamma )\sin x_{ji}},\ \ \ \ \
(i\neq j), \nonumber \\
{\mathcal{B}}_{ij} &=&e^{-p_{j}}b_{j}^{^{\prime }}\frac{\sin
(x_{i}+x_{j}+\lambda )}{\sin \lambda \sin (x_{i}+x_{j})}-e^{-p_{i}}b_{j}%
\frac{\sin (x_{i}+x_{j}+\lambda +4\gamma )}{\sin (\lambda +4\gamma )\sin
(x_{i}+x_{j})},	 \nonumber \\
\ \ \ {\mathcal{C}}_{ij} &=&e^{p_{j}}b_{j}\frac{\sin (x_{i}+x_{j}-\lambda )}{%
\sin \lambda \sin (x_{i}+x_{j})}-e^{p_{i}}b_{j}^{^{\prime }}\frac{\sin
(x_{i}+x_{j}-\lambda -4\gamma )}{\sin (\lambda +4\gamma )\sin (x_{i}+x_{j})},
\nonumber \\
{\mathcal{A}}_{ii} &=&(\cot (\gamma )+\cot (\lambda ))e^{p_{i}}b_{i}-(\cot
(\lambda +\gamma )-\cot (\gamma ))e^{-p_{i}}b_{i}^{^{\prime }}	\nonumber \\
&&+\sum_{k\neq i}^{2}\left( (\cot (x_{ik}+\gamma )-\cot
(x_{ik})e^{p_{k}}b_{k}\right)  \nonumber \\
&&+\frac{\sin (x_{ik}+\lambda +4\gamma )\sin (x_{ki}+\lambda +\gamma )\sin
\gamma }{\sin (\lambda +4\gamma )\sin x_{ik}\sin (x_{ki}+\gamma )\sin
(\lambda +\gamma )}e^{-p_{i}}b_{k}^{^{\prime }}) \nonumber \\
&&+\sum_{k=1}^{2}((\cot (x_{i}+x_{k}+\gamma )-\cot
(x_{i}+x_{k})e^{-p_{k}}b_{k}^{^{\prime }} \nonumber \\
&&+\frac{\sin (x_{i}+x_{k}+\lambda +4\gamma )\sin (x_{i}+x_{k}-\lambda
-\gamma )\sin \gamma }{\sin (\lambda +4\gamma )\sin (x_{i}+x_{k})\sin
(x_{i}+x_{k}-\gamma )\sin (\lambda +\gamma )}e^{-p_{i}}b_{k}) \nonumber \\
{\mathcal{D}}_{ii} &=&(\cot (\gamma )+\cot (\lambda
))e^{-p_{i}}b_{i}^{^{\prime }}-(\cot (\lambda +\gamma )-\cot (\gamma
))e^{p_{i}}b_{i} \nonumber \\
&&+\sum_{k\neq i}^{2}\left( (\cot (x_{ki}+\gamma )-\cot
(x_{ki})e^{-p_{k}}b_{k}^{^{\prime }}\right)  \nonumber \\
&&+\frac{\sin (x_{ki}+\lambda +4\gamma )\sin (x_{ik}+\lambda +\gamma )\sin
\gamma }{\sin (\lambda +4\gamma )\sin x_{ki}\sin (x_{ik}+\gamma )\sin
(\lambda +\gamma )}e^{p_{i}}b_{k}) \nonumber \\
&&+\sum_{k=1}^{2}((\cot (x_{i}+x_{k})-\cot (x_{i}+x_{k}-\gamma
))e^{p_{k}}b_{k} \nonumber \\
&&+\frac{\sin (x_{i}+x_{k}-\lambda -4\gamma )\sin (x_{i}+x_{k}+\lambda
+\gamma )\sin \gamma }{\sin (\lambda +4\gamma )\sin (x_{i}+x_{k})\sin
(x_{i}+x_{k}+\gamma )\sin (\lambda +\gamma )}e^{p_{i}}b_{k}^{^{\prime
}}).\label{me2}
\end{eqnarray}
The $L,M$ satisfies the Lax equation Eq.(\ref{laxeq})
and the Hamiltonian $H$ can also be rewritten in the form of
Eq.(\ref{Ham}).

The function-independent Hamiltonian flows  can be generated by	  calculating
the characteristic polynomial of that Lax matrix $L$
\begin{equation}
\det (L-v\cdot Id)=\sum_{j=0}^{4}\frac{(\sin \lambda )^{^{(j-1)}}\sin
(\lambda +j\gamma )}{(\sin (\gamma +\lambda
))^{j}}(-v)^{4-j}H_{j},
\end{equation}
where $H_{0}=H_{4}=1$, $H_{1}=H_{3}=H$.
$H$ and $H_{2}$ have the same forms as Eq.(\ref{h1}) and
(\ref{h2}).

\vspace{0.7cm}

\noindent{\bf Remarks:}
\begin{enumerate}
  \item As far as the forms   of the Lax pair for the
rational-type  systems	are concerned, we
can get	 them	by making the following
substitutions
\begin{eqnarray*}
\sin x &\rightarrow &x, \\
\cos x &\rightarrow &1,
\end{eqnarray*}
for all the above statements.

\item  It should be pointed out	 that the Lax pair given
in Eqs.(\ref{lax})-(\ref{me1}) which  are without spectral
parameter can be   derived from the one with spectral
parameters (see Eqs.(\ref{l2})-(\ref{me2})) by taking the following limit

\[
\lambda \rightarrow i\infty ,
\]
up to an appropriate gauge transformation of the Lax matrix with a diagonal
matrix.
\end{enumerate}

\section{Summary and discussions}

In this paper, we propose the Lax pairs for
trigonometric $C_{2}$ RS model together with its rational limit
and show their integrability. Involutive Hamiltonians are shown to
be generated by the characteristic polynomial of the Lax matrix.
In the nonrelativistic limit, the system leads to CM system associated with the
root system of $C_{2}$ which is known previously. It is expected  that,
in the general case of $C_{n}$ for $n\geq 2$, the explicit expressions of $L$
and $M$ must have similar forms as those presented here.
So it would be	interesting to make some progress in this respect
in the very near future.

\section*{Acknowledgement}

One of the authors K. Chen  is grateful to professors K. J. Shi and L. Zhao for
their encouragement.


\vspace{5pt}


\begin{thebibliography}{99}

\bibitem{h1}  Hasegawa K 1997 {\sl Commun. Math. Phys.} {\bf 187} 289

\bibitem{nksr} Nijhoff F W, Kuznetsov V B, Sklyanin  E K  and  Ragnisco O
1996 {\sl J. Phys.} {\bf A29} L333


\bibitem{gm}  Gorsky A and  Marshakov A 1996 {\sl Phys. Lett.} {\bf B375} 127

\bibitem{n}  Nekrasov N 1998 {\sl Nucl. Phys.} {\bf B531}  323

\bibitem{bmmm}	Braden H W, Marshakov A, Mironov A and Morozov A
1999 {\sl Nucl. Phys.} {\bf B558} 371

\bibitem{op}  Olshanetsky M A and Perelomov A M 1981 {\sl Phys. Rep.} {\bf 71} 314

\bibitem{in} Inozemtsev V I 1989 {\sl Lett. Math. Phys.} {\bf 17}  11

\bibitem{hp1}	D'Hoker E and Phong D H	 1998 {\sl Nucl. Phys.} {\bf B530} 537

\bibitem{bcs}  Bordner A J, Corrigan E and  Sasaki R 1998
{\sl Prog. Theor. Phys.} {\bf 100}  1107


\bibitem{hm} Hurtubise J C and	Markman E {\sl e-print} {\tt  math/9912161}

\bibitem{ko1}  Komori Y and Hikami K 1998 {\sl J. Math. Phys.} {\bf 39} 6175

\bibitem{ko2}  Komori Y {\sl e-print} {\tt  math.QA/9910003}

\bibitem{di}  Diejen J F van 1994 {\sl J. Math. Phys.} {\bf 35} 2983

\bibitem{di1}  Diejen J F van  1995 {\sl Compositio. Math.} {\bf 95}  183

\bibitem{h2}  Hasegawa K,  Ikeda T  and	  Kikuchi T 1999  {\sl J. Math.
Phys.} {\bf 40} 4549


\bibitem{kai1}	 Chen K,  Hou B Y, Yang W L and Zhen Y 1999
{\sl Chin. Phys. Lett.} {\bf 16} 1; 1999 {\sl High energy physics and nuclear
physics.} {\bf 23}  854

\bibitem{kai2}	Chen K, Fan H,	Hou B Y, Shi K J,  Yang W L and
Yue R H 1999 {\sl Prog. Theor. Phys. Suppl.} {\bf 135} 149

\bibitem{ki}  Kikuchi T {\sl e-print} {\tt math/9912114}

\bibitem{bcs1}	Bordner A J, Corrigan E and Sasaki R 1999
{\sl Prog. Theor. Phys.} {\bf 102}  499

\bibitem{r1}  Ruijsenaars  S N M  1987 {\sl Comm. Math. Phys.} {\bf 110}  191

\bibitem{bc} Bruschi M and Calogero F 1987 {\sl Commun. Math. Phys.} {\bf
109} 481

\bibitem{kz}   Krichever  I and Zabrodin A 1995	 {\sl Usp. Math. Nauk} {\bf 50:6} 3

\bibitem{s1}  Suris Y B {\sl e-print} {\tt hep-th/9602160}

\bibitem{s2}  Suris Y B	 1997  {\sl Phys. Lett.} {\bf A225} 253

\bibitem{r2} Ruijsenaars   S N M 1988 {\sl Comm. Math. Phys.} {\bf 115} 127

\end{thebibliography}
\end{document}